\documentclass[pdflatex,sn-nature]{sn-jnl}

\usepackage{graphicx}%
\usepackage{pdfpages}
\usepackage{multirow}%
\usepackage{amsmath,amssymb,amsfonts}%
\usepackage{amsthm}%
\usepackage{mathrsfs}%
\usepackage{xcolor}%
\usepackage{textcomp}%
\usepackage{booktabs}%
\usepackage{algorithm}%
\usepackage{algorithmicx}%
\usepackage{algpseudocode}%
\usepackage{listings}%
\usepackage[section]{placeins}%

\theoremstyle{thmstyleone}%
\theoremstyle{thmstyletwo}%

\theoremstyle{thmstylethree}%

\begin{document}

\title[Paradigm Choice in EEG Biometric Security]{Adversarial Vulnerabilities of Neural Biomarker Identification Systems}


\author*[1,2]{\fnm{Polina} \sur{Tapal}}\email{polina.tapal.001@student.uni.lu}

\author[1,3]{\fnm{Bryce-Allen} \sur{Bagley}}

\affil[1]{\orgname{Cerberus Neurosecurity Research Institute (CNRI)}, \orgaddress{\city{San Francisco}, \state{CA}, \postcode{94102}, \country{USA}}}

\affil[2]{\orgdiv{Department of Computer Science}, \orgname{University of Luxembourg}, \orgaddress{\street{435 Av. du Rock'n'Roll}, \city{Esch-sur-Alzette}, \postcode{4361}, \country{Luxembourg}}}

\affil[3]{\orgname{Cognitive Security Institute}}

\abstract{There is growing interest in the proposed use of EEG signals as biometric credentials, but thus far there has been little research on the reliability and security of such biometrics. Prior adversarial tests have focused on deep-learning classifiers and assumed attackers have full access to the classifier model. This has left unexamined other, more popular categories of neural signature methods as well as the more realistic case of an adversary having only black-box access to a classifier. In this paper we develop a collection of adaptive attack algorithms which learn to fool an authentication system via targeted alterations of stolen EEG recordings, without requiring any knowledge of the authentication system itself. Tested on 6 public datasets spanning three recording conditions (reacting to visual stimuli, imagining hand movements, and resting), it reveals that different signaturing approaches vary significantly in their degrees of vulnerability to adversarial attacks. We show that vulnerability to spoofing attack is greatly impacted by the recording conditions, with significant variation depending on task at time of recording. Finally, we provide recommendations for improving neural signature biometrics based on the results of our adversarial testing.}

\maketitle

\section{Introduction}\label{sec1}

The convergence of neuroscience and security, termed \emph{neurosecurity}, was first articulated by Denning, Matsuoka and Kohno~\cite{denning2009} in the context of implantable neural devices, and has since expanded significantly as non-invasive interfaces based on technologies such as Electroencephalography (EEG)  have moved from laboratory settings into consumer and clinical deployment. Early framing by Ienca and Haselager~\cite{ienca2016hacking} identified brain--computer interfaces (BCIs) as a surface for both offensive exploitation and privacy violation. Their subsequent work with Emanuel~\cite{ienca2018} extended these concerns to consumer neurotechnology, characterising the risk of involuntary neural data disclosure as brain leaks. Pycroft et al.~\cite{pycroft2016} coined the term brainjacking for adversarial manipulation of implanted stimulators, while Canham and Sawyer~\cite{canham2020masint} argued that passively captured neural emissions constitute a new category of measurement and signals intelligence (MASINT), placing the human brain within the threat landscape of intelligence collection. Further work has demonstrated concrete attack vectors and characterised the BCI attack surface across multiple architectures~\cite{lange2018sidechannel, bernal2020implants, bonaci2014exocortex, landau_mind, belkacemp300, bernal2021survey, bernal2023eight}. The most recent and comprehensive survey of this landscape is Bagley et al.~\cite{bagley2026cerberus}, who catalogue threat vectors and state-of-the-art defensive methods across neurotechnology broadly. Collectively, this body of work establishes that neural data are sensitive, leakable, and increasingly targetable, motivating the design of authentication systems whose security properties are analysed under explicit adversarial models.

A parallel strand of research has investigated \emph{neural signatures}: the stable, subject-specific features of brain activity that can serve as biometric identifiers. Campisi and La Rocca~\cite{campisi2014brain} provided an early systematic treatment of EEG-based biometric recognition, establishing that individual differences in spectral and event-related features are sufficiently stable for identity verification. Ruiz-Blondet, Jin and Laszlo~\cite{ruizblondet2016cerebre} subsequently demonstrated near-perfect identification accuracy using the averaged ERP in their CEREBRE protocol, confirming the richness of individually distinctive neural response patterns. Barachant et al.~\cite{barachant2013riemannian} demonstrated that spatial covariance matrices of EEG recordings, treated as points on the Riemannian manifold of symmetric positive definite (SPD) matrices, carry highly differentiating identity information for BCI classification, establishing the geometric foundation that one of our authenticators directly inherits. Subsequent work formalized EEG-based person identification as a biometric problem. Wang et al.~\cite{wang2020brainprint} showed that resting-state connectivity graphs yield stable brainprints across sessions, and Zhang et al.~\cite{zhang2021review} survey the state of EEG authentication methods, cataloging both event-related potential (ERP) and resting-state paradigms. On the attack side, Bernal et al.~\cite{bernal2022jamming} demonstrated neuronal jamming cyberattacks that corrupt BCI task performance via crafted stimuli, Wu et al.~\cite{wu2022adversarial} provide a systematic review of adversarial attacks and defenses across physiological computing modalities including EEG, and Yi et al.~\cite{yi2024timefrequency} demonstrate imperceptible time-frequency adversarial attacks against deep-learning brainprint recognizers specifically, though requiring gradient access to the target model. Taken together, the literature shows that neural signatures are both powerful biometric primitives and non-trivial attack targets. 

The present paper contributes an adversarial evaluation framework for geometry-based EEG authentication. For our primary attack algorithm, we treat perturbation strategy selection as a multi-armed bandit problem, enabling an adaptive black-box adversary (requiring no model access beyond accept/reject feedback) to discover which families of signal-level attack are most effective against nearest-centroid and Gaussian Na\"ive Bayes classifiers operating on the same covariance manifold features, without requiring prior knowledge of the recording paradigm used to collect the EEG data.

\section{Methods}\label{sec2}

\subsection{Threat model and scope}\label{subsec2-1}

We consider an adversary attempting to defeat an EEG-based biometric authentication system by submitting a forged neural signal. The system authenticates users by verifying that a presented EEG trial is consistent with an enrolled subject's covariance profile, summarized as a per-identity template $\boldsymbol{\mu}_k$ constructed at enrollment and defined in \S\ref{subsubsec2-2-2}. The adversary's goal is to produce a signal that is accepted by the system under a false claimed identity, that is, to maximize the False Acceptance Rate (FAR). 

The signal being perturbed is a raw multichannel EEG trial $X \in \mathbb{R}^{C \times T} $, recorded from $C$ electrodes over $T$ time samples. The adversary has black-box access to the authenticator, submitting trials and observing accept/reject decisions, with no access to the enrolled templates or the decision threshold $\tau$. The concrete case under consideration is one in which the attacker has obtained a prior recording, for instance, through a compromised acquisition device or access to a previously captured session. The requirement is weaker than that scenario suggests, and we state it in its weakest form. Because the feature map $\phi$ discards the temporal axis, the trial at hand needs to match the deployed system only in channel count and montage, and not in trial length, recording session, or recording paradigm. It does not need to be a recording of the identity under attack, nor of any enrolled subject at all. We accordingly call it the donor trial and reserve target identity for the enrolled identity $k'$ under which acceptance is sought. We note that the party harmed is the target identity, while the donor may be any subject, enrolled or not. \S\ref{subsec3-4} tests this limit directly by drawing donors from a different recording paradigm than the one used at enrollment.

The authenticator operates entirely on the log-covariance vector $\mathbf{v} = \phi(X)$ and never on the raw signal, so any perturbation that displaces the covariance geometry of $X$ toward $\boldsymbol{\mu}_{k'}$ is, from the system's view, indistinguishable from a genuine trial recorded from subject $k'$. Conversely, perturbations that do not affect covariance structure, such as a uniform DC offset, have no effect on the authenticator whatsoever. This motivates studying perturbations at the signal level, before feature extraction, rather than restricting the adversary to manipulating the feature vector directly. The attacks studied here are impersonation attacks. The adversary submits a perturbed donor trial under claimed identity $k'$ and succeeds when the system accepts it.

The assumption that the adversary holds a prior recording rather than synthesizing EEG from scratch is grounded in the current limitations of EEG generation. Unlike images or audio, realistic full-synthesis of subject-specific EEG signals that preserves fine-grained covariance structure remains an open problem~\cite{gan}. Replay and perturbation-based attacks are therefore the most credible threat vector in practice. Rather than focusing on a single fixed attack, we evaluate a trio of strategies: a stochastic multi-armed bandit (\S\ref{subsec2-3}), a black-box VAE trained via REINFORCE, and a gray-box surrogate-gradient GAN with true backpropagation (\S\ref{subsubsec2-3-1}).

\subsection{Base authentication system}\label{subsec2-2}

\subsubsection{Feature extraction}\label{subsubsec2-2-1}

Let a single EEG trial be a matrix $X \in \mathbb{R}^{C \times T}$, where $C$ is the number of channels and $T$ is the number of time samples. Its empirical spatial covariance is
\begin{equation}
\mathbf{Cov} = \frac{1}{T-1}XX^T + \epsilon I_C, \label{eq:cov}
\end{equation}
with $\epsilon = 10^{-6}$ ensuring positive definiteness. We compute $XX^T$ rather than $X^TX$ because we are interested in spatial, not temporal, relationships, yielding a $C \times C$ matrix on the symmetric positive definite (SPD) manifold. This manifold is curved under the natural affine-invariant metric~\cite{thanwerdas2021oinvariant} (Supplementary Note 2), so we linearize it via the Log-Euclidean map~\cite{arsigny2007logeuclid}. Writing the eigenvalue decomposition $\mathbf{Cov}=U\Lambda U^T$, $S = \log_m(\mathbf{Cov}) = U\,\mathrm{diag}(\log\lambda_1,\dots,\log\lambda_C)\,U^T$ sends the trial to the tangent space at the identity, a flat space where Euclidean operations are geometrically meaningful. The upper triangular entries of $S$ (the lower triangle is redundant by symmetry) are then extracted and $\ell_2$-normalized to yield a feature vector $v \in \mathbb{R}^D$, where $D = C(C+1)/2$. The normalization is deliberate. The magnitude of $\mathbf{v}$ encodes non-stationary signal power (impedance drift, fatigue), while its direction encodes the stable, participant-specific covariance structure. Discarding magnitude and mapping every trial onto the unit sphere $\mathbb{S}^{D-1}$ makes cosine similarity the natural verification metric (\S\ref{subsubsec2-2-2}). 

To confirm this log-map is not incidental, we compare it against two feature ablations across all eight evaluation streams, an Euclidean covariance feature (same extraction, no log-map) and a per-channel power-spectral-density feature in Supplementary Note 3. On RSVP and motor imagery, both baselines collapse to near-chance discrimination, with equal error rate (EER, the operating point at which false acceptance and false rejection rates coincide) of $0.31$--$0.44$, whereas Riemannian features achieve $0.07$--$0.15$. On resting-state, all three features are comparably poor ($\approx 0.31$--$0.50$), confirming that resting-state vulnerability is structural, not feature-choice-dependent.

\subsubsection{Classifiers: nearest-centroid and Gaussian Na\"ive Bayes}\label{subsubsec2-2-2}

We evaluate the same log-covariance features $\mathbf{v}\in\mathbb{R}^D$ under two classifiers with near-opposite structural assumptions, so that agreement between them is informative about whether a finding reflects the feature geometry or one decision rule built on top of it. 

\paragraph{Nearest-centroid.} Let $\{(\mathbf{v}_i, y_i)\}_{i=1}^{N}$ be the enrollment set. For each identity $k$ we average the normalized log-covariance vectors of that participant's enrollment trials and renormalize, giving the template $\boldsymbol{\mu}_k = \bar{\mathbf{v}}_k / \|\bar{\mathbf{v}}_k\|$. To verify a claimed identity $k'$, a query $q$ is scored by cosine similarity $\mathrm{sim}(q, \boldsymbol{\mu}_{k'})$, which the $\ell_2$-normalization of \S\ref{subsubsec2-2-1} makes the natural inner product on the unit sphere. The explicit
formula, and a comparison with the deep-learning alternative, are given in Supplementary Note 2. Nearest-centroid (NC) is deliberately simple, as a learned decision boundary would conflate the adversarial surface of the classifier with that of the feature space, whereas nearest-centroid attributes all attack effectiveness to displacement in SPD geometry -- an assumption we test directly below with a classifier that does have a learned boundary.

\paragraph{Gaussian Na\"ive Bayes.} For each enrolled identity $k$, we instead fit a dedicated one-vs-all Gaussian Na\"ive Bayes (GNB) classifier $g_k$ on the same vectors, treating $k$'s trials as the positive class and all other identities' trials as the negative~\cite{arias_n400faces}. GNB models each feature dimension as an independent univariate Gaussian conditioned on class. This is an axis-aligned assumption, in contrast to nearest-centroid, which scores along a single global direction. The verification score is $P(\text{genuine} \mid q) = g_k(q)$ in place of $\mathrm{sim}(q,\boldsymbol{\mu}_k)$. Threshold calibration, the UCB bandit, the attack arms, and the dual reward are otherwise unchanged from \S\ref{subsec2-3}--\S\ref{subsec2-4}. Results for both classifiers are reported side by side throughout \S\ref{sec3}.

\paragraph{Linear support vector machine.} To test whether the pattern observed under nearest-centroid and GNB depends specifically on using classifiers with no learned decision-boundary capacity, we additionally repeat the full protocol with a one-vs-all linear SVM per enrolled identity. This classifier optimizes a decision boundary rather than relying on fixed geometric or probabilistic structure. Full methodological detail, including a threshold-calibration procedure this classifier's learning capacity requires, is given in Supplementary Note 9; results are reported alongside NC and GNB in \S\ref{subsec3-3}.

\subsubsection{Threshold calibration}\label{subsubsec2-2-3}

Setting $\tau$ at the $p$-th percentile of genuine scores bounds the FRR at $p/100$ on the training population:
\begin{equation}
\mathrm{FRR} \leq \frac{p}{100}. \label{eq:frr}
\end{equation}
To select $p$, we examined the EER-crossing percentile $p_{\mathrm{EER}}$ under the nearest-centroid classifier across all eight evaluation streams (Table~\ref{tab:baseline_summary}; full breakdown in Supplementary Tables 4--6). The two RSVP datasets ($p_{\mathrm{EER}}=6$ for Zhang, $p_{\mathrm{EER}}=8$ for Won) and the two motor imagery datasets ($p_{\mathrm{EER}}=4$ for Cho, $p_{\mathrm{EER}}=8$ for Lee) cluster tightly around $p\approx10$, while the four resting-state streams cross at substantially higher percentiles ($p_{\mathrm{EER}} \in \{34,36,38,50\}$; \S\ref{sec4}). We use $p \approx 10$ as a reference operating point in the experiments that follow, as it sits within or immediately adjacent to the empirical EER band for four of the eight streams and gives a single, dataset-independent percentile for cross-paradigm comparison. However, since the optimal balance of FAR and FRR is application-dependent, the adversarial evaluation in Section~\ref{subsec2-3} reports results across a range of $p$ values, allowing the authenticator's resilience to be characterized as a function of the operating point rather than at a single fixed threshold.

\subsection{Adversarial evaluation}\label{subsec2-3}

Score-feedback black-box attacks on EEG biometrics were first demonstrated via hill-climbing optimization of synthetic templates~\cite{maiorana2013hillclimbing}. We instead frame strategy selection as a bandit over interpretable, signal-level perturbation families, which additionally exposes which attack family is effective per paradigm, rather than only whether an attack succeeds. We frame the adversarial search as a stochastic multi-armed bandit problem. Let $A = \left\{a_0, \dots, a_7\right\}$ be a set of 8 attack arms: signal perturbations applied to a raw EEG trial to produce a forgery. The attacker pursues a dual objective: maximizing FAR, while minimizing perturbation magnitude, so that forgeries are both effective and minimally distorted relative to the original signal. 

At each round $t$, the UCB policy selects the arm:
\begin{equation}
a^*(t) = \operatorname{argmax}_{a \in A}\left[\hat{\mu}_a(t) + c\sqrt{\frac{\ln t}{N_a(t)}}\right] \label{eq:ucb}
\end{equation} where $\hat{\mu}_a(t) = R_a(t)/N_a(t)$ is the empirical mean reward for arm $a$, $N_a$ is its pull count and $c = \sqrt{2}$ is the exploration constant. The reward signal is a scalarization of two objectives:
\begin{equation}
r_t = (1 - \lambda)\cdot\mathbf{1}\!\left[\mathrm{sim}(q,\,\boldsymbol{\mu}_{k'}) \geq \tau\right] \;+\; \lambda\cdot\left(1 - \frac{\|X' - X\|_F}{\|X\|_F}\right) \label{eq:reward}
\end{equation}
where $\lambda \in [0,1]$ is a trade-off parameter. The first term rewards false acceptance, and the second rewards perturbation minimality, normalized to $[0,1]$. Setting $\lambda = 0$ recovers the purely binary objective. Larger $\lambda$ penalizes arms that achieve acceptance only through large signal distortions. We refer to $\|X' - X\|_F / \|X\|_F$ as the attack budget consumed by a given arm on a given trial. It measures the fraction of the original signal energy added by the perturbation. Arms that cross the acceptance threshold only by consuming a large fraction of this budget receive lower reward under $\lambda > 0$, and the bandit learns to favor arms that achieve false acceptance at low budget cost. We fix $\lambda = 0.3$ for all experiments reported. Supplementary Note 5 sweeps $\lambda \in \{0, 0.1, 0.3, 0.5\}$ and shows adversarial FAR changes by at most 0.073 across this range, with the paradigm-specific arm-family finding unchanged at every value, so 0.3 is representative of the swept range rather than a value the result depends on. The UCB bandit thus selects arms that jointly maximize attack effectiveness and forgery quality, rather than rewarding any perturbation that happens to cross the threshold. Since this formulation is expressed in terms of nearest-centroid's cosine similarity, we also specify its counterpart for GNB. In this case $g_k(q)$ substitutes for $\mathrm{sim}(q,\boldsymbol{\mu}_{k'})$ (\S\ref{subsubsec2-2-2}) with no other change to the bandit, arms, or reward.

Each arm applies a distinct signal-level perturbation to the new raw trial $X \in \mathbb{R}^{C \times T}$ (see Table~\ref{tab:perturbations}).

\begin{table}[h]
\caption{Signal-level perturbation arms applied to raw EEG trials $X \in \mathbb{R}^{C \times T}$. Arm $a_3$ operates on the extracted, already-bandpassed trials, not raw signal. RSVP streams are filtered to 1--10\,Hz, so only $\delta$ and part of $\theta$ carry any content there; the full $\delta$--$\beta$ range is only available on the resting-state and motor-imagery streams (1--30\,Hz). $\gamma$ is listed for completeness of the arm's general definition; no stream retains content above 30\,Hz, so $\gamma$-band gain is a no-op for all eight streams.}\label{tab:perturbations}
\begin{tabular}{@{}cp{3.5cm}p{7cm}@{}}
\toprule
\textbf{Arm} & \textbf{Name} & \textbf{Perturbation} \\
\midrule
$a_0$ & Gaussian noise & $X' = X + \mathcal{N}(0,\sigma^2 I)$ \\
$a_1$ & Correlated noise & $X'= X + \sigma Dw$, structured via random subspace \\
$a_2$ & Temporal distortion & segment shuffle, crop-pad, or cyclic shift \\
$a_3$ & Frequency manipulation & bandpass-selective gain in $\delta, \theta, \alpha, \beta, \gamma$ bands \\
$a_4$ & Covariance interpolation & $X'=(1 - \alpha)X + \alpha X_{\mathrm{imp}}$, $\alpha \in [0,0.5]$ \\
$a_5$ & Replay + jitter & $X' = X + \varepsilon + \mathrm{roll}(X,s)$ \\
$a_6$ & Channel coupling & $X_i' \leftarrow X_i + \alpha X_j$ for a random pair $(i,j)$ \\
$a_7$ & Eigenvalue perturbation & reshape covariance spectrum via amplify/flatten \\
\botrule
\end{tabular}
\end{table}

The eight arms were selected to provide coverage across qualitatively distinct attack families, each targeting a different property of the Riemannian feature pipeline: additive noise ($a_0$, $a_1$), temporal rearrangement ($a_2$), spectral gain ($a_3$), signal space interpolation ($a_4$), replay ($a_5$), channel coupling ($a_6$), and eigenvalue reshaping ($a_7$). The bandit is then left to discover empirically which family is most effective, rather than assuming a single dominant attack vector. The arm set deliberately contains no null perturbation: every arm modifies the trial. On a paradigm where an untouched impostor recording is already accepted most of the time, the bandit cannot do better than the baseline. Negative attacker lift (\S\ref{subsec3-2}) therefore reports that no perturbation family improves on inaction. A real attacker facing such a system would not perturb it at all.

\subsubsection{Generative attackers: VAE and surrogate-gradient GAN}\label{subsubsec2-3-1}

As an alternative to the UCB bandit, we evaluate two generative attackers under the identical dual reward, $N_{\mathrm{iter}}=1{,}000$-iteration budget, and enrollment/test splits. A VAE-attacker~\cite{kingma2014vae} (black-box, trained online via REINFORCE~\cite{williams1992reinforce}) encodes a flattened victim trial to a posterior over a 32-dimensional latent space and decodes a perturbation $\delta$, exploring via the posterior's stochasticity; full architecture in Supplementary Note 4. A surrogate-gradient GAN (SGAN) escalates to a gray-box threat model. Differentiating the deployed feature pipeline would require white-box access. The attacker instead trains a small differentiable surrogate $S_\theta$ on enrollment data to approximate the verification score, then trains a generator and discriminator against it with true backpropagation, following the surrogate-attack paradigm of Papernot et al.~\cite{papernot2017practical} with a Carlini--Wagner-style soft perturbation-norm penalty~\cite{carlini2017evaluating} in place of a hard constraint. Full loss, architecture, and diagram are in Supplementary Note 4. FAR is measured on the nearest-centroid authenticator throughout for comparability across all three attackers. The full attacker $\times$ classifier grid, including both generative attackers against GNB and the linear SVM, is reported in Supplementary Note 10.

\subsection{Experimental protocol}\label{subsec2-4}

Experiments are repeated across a range of threshold percentiles $p \in \{5, 10, 15, 20, 25\}$ to characterise how authenticator resilience varies with the FAR/FRR operating point. All trials are drawn from a single participant $\times$ day session. With each session, a fraction $\rho = 0.5$ of trials per subject are reserved for enrollment (i.e.\ fitting the per-class template or model -- $\mu_k$ for nearest-centroid, $g_k$ for GNB -- and calibrating $\tau$). The remaining trials form the test pool from which the attack samples donor signals $X$. A two-pool split suffices because $p$ is chosen on a system-design criterion independent of adversarial outcomes, so no feedback loop between the attack evaluation and the threshold selection exists. A total of $N_{\mathrm{iter}} = 1000$ UCB iterations are run per day. This is repeated independently across days and subjects.

To assess whether arm selection is non-trivial, we apply the one-sided Mann-Whitney~U test (Wilcoxon rank-sum) between the reward sequences of the best-performing and worst-performing arms within each dataset, pooled across
days.  Let $\{r^{(b)}_t\}$ and $\{r^{(w)}_t\}$ denote the rewards collected across all iterations in which arm $b$ (best) and arm $w$ (worst) were pulled respectively; the sequences have unequal length because the UCB policy allocates pulls unevenly. The null hypothesis is that the two distributions are identical. Rejection at $\alpha < 0.05$ confirms that the bandit has identified a statistically meaningful difference in attack effectiveness. Results are reported in Supplementary Table 3.

We additionally test the montage-compatibility claim of \S\ref{subsec2-1} directly. For each ordered pair of the eight streams, the authenticator is enrolled on one and donor trials are drawn exclusively from the other, with the bandit, arm set, and reward unchanged. All streams are reduced to the 31 electrodes they share ($C=31$, $D=496$). Donors are truncated to the enrolled trial length, band-matched to the enrolled passband, and rescaled to the enrolled amplitude. The rescaling is necessary because the Log-Euclidean map is not scale-invariant, a global factor $s$ contributing $\log(s)I$
that survives the $\ell_2$ normalization, and matching gain is a capability any adversary possesses. Where band-matching is required, the identical filter is applied to the enrolled trials, so that neither side passes through more filtering than the other. Because this harmonization alters the montage, the diagonal of the resulting grid rather than Table~\ref{tab:adv_summary} is the within-condition reference. The full grid is in Supplementary Note 11. 

\subsection{Datasets}\label{subsec2-5}

Prior to the dataset description, we fix terminology that may not be self-evident to readers without an EEG background. A \textbf{trial} is a single segment of multichannel EEG, represented as a matrix $X \in \mathbb{R}^{C \times T}$. In event-related paradigms a trial is a stimulus-locked epoch, a window extracted around a known stimulus onset and spanning the expected response interval. In resting-state paradigms there are no stimuli, so trials are instead fixed-length non-overlapping windows cut from a continuous recording. A \textbf{session} is a single recording sitting, one participant on one day, yielding some number of trials.

We evaluate the adversarial framework across three paired experiments, each representing a qualitatively distinct category of neural driver: (i) exogenous stimulation, in which discriminative brain responses are time-locked to external stimuli presented by the system, (ii) no neural driver, in which EEG is collected at rest without any task or stimulus structure, (iii) endogenous cognitive effort, in which discriminative neural activity is generated internally by the  participant's voluntary mental imagery. This taxonomy allows us to disentangle the contribution of the neural driver to both baseline biometric discriminability and adversarial vulnerability, as well as to test whether the bandit's arm-preference structure is paradigm-specific or generic across signal types. Zhang and Wu~\cite{zhang2019vulnerability} priorly showed CNN task decoders are vulnerable to gradient-based attack across P300, error-related negativity, and motor-imagery paradigms; our taxonomy instead targets nearest-centroid and Gaussian Na\"ive Bayes biometric authenticators built on the same Riemannian covariance features, rather than task decoders, and asks whether paradigm-level vulnerability is a property of the classifier or of the covariance geometry itself. 

The first two datasets (Zhang et al.\ and Won et al.) employ the Rapid Serial Visual Presentation (RSVP) oddball paradigm, in which brief stimuli are displayed at a fixed rate and the subject attends to rare target items. This paradigm reliably elicits the P300 event-related potential, a positive deflection at parietal and occipital electrodes peaking 300–600\,ms post-stimulus, whose amplitude, latency, and spatial distribution carry individually distinctive patterns that have been exploited for ERP-based biometric identification~\cite{campisi2014brain}. We chose RSVP because (i) the P300 component is the most studied EEG biometric signature, (ii) two publicly available datasets with compatible recording setups both use this paradigm, enabling a direct cross-dataset comparison without paradigm confounds, and (iii) the within-session enrollment/verification protocol we adopt is the standard evaluation setting in P300 biometrics.

Two further datasets provide resting-state EEG with no stimulus structure. Participants either fixate a central point (eyes open, EO) or sit quietly with eyes closed (EC) for a fixed duration. The EO and EC conditions of each dataset are treated as fully independent evaluations, yielding four resting-state evaluation streams in total. CP5 is not recorded in all COG-BCI participants and is dropped globally from both resting-state datasets, giving $C=31$ and $D=496$ on these four streams. Wang et al.\ provides long inter-session gaps (90\,min and one month); Hinss et al.\ provides a weekly test-retest interval. Together they probe whether the Riemannian covariance signature remains stable across qualitatively different retest timescales, and whether the adversarial arm preferences observed under ERP paradigms replicate in the absence of any stimulus-locked neural response. 

Two final datasets (Cho et al.\ and Lee et al.) provide motor imagery EEG. Participants imagine left- or right-hand movement, generating mu (8--12\,Hz) and beta (13--30\,Hz) event-related desynchronisation that is spatially localised over contralateral sensorimotor cortex and individually distinctive. Left- and right-hand trials are pooled per subject; a wider bandpass of 1--30\,Hz is used to retain the physiologically relevant mu/beta bands, and channel selection is restricted to the same 32-electrode Won subset, yielding identical feature dimensionality ($C=32$, $D=528$) to the RSVP pair.

Full recording parameters and preprocessing pipelines are given in Supplementary Note 6.

\subsection{Ethical considerations}\label{subsec2-6}

The study reanalyses six previously published, publicly available EEG datasets~\cite{zhang_dataset,won_dataset,wang2022testretest,hinss2023cogbci,cho2017mi,lee2019mi}. No new human-subjects data were collected by the present authors. Each dataset was collected under the ethical approval and informed-consent procedures reported by its original publication (Supplementary Note 6). All recordings used here are de-identified and released for secondary research use.

\section{Results}\label{sec3}

\subsection{Baseline authentication performance}\label{subsec3-1}

\begin{table}[h]
\caption{Baseline authentication performance across all eight streams (nearest-centroid classifier; GNB baseline reported separately in Table~\ref{tab:gnb_summary}). $p_{\mathrm{EER}}$: threshold percentile at EER crossing; FAR at $p{=}10$: non-adversarial false acceptance rate at the reference operating point ($N=15$ for RSVP, $N=20$ otherwise).}\label{tab:baseline_summary}
\begin{tabular}{@{}llccc@{}}
\toprule
Paradigm & Dataset & $p_{\mathrm{EER}}$ & EER & FAR ($p{=}10$) \\
\midrule
\multirow{2}{*}{RSVP}
  & Zhang      &  6 & 0.154 & 0.103 \\
  & Won        &  8 & 0.153 & 0.111 \\
\midrule
\multirow{4}{*}{Resting-state}
  & Wang-EO    & 38 & 0.389 & 0.818 \\
  & Wang-EC    & 34 & 0.345 & 0.812 \\
  & COG-BCI-EO & 50 & 0.492 & 0.896 \\
  & COG-BCI-EC & 36 & 0.391 & 0.842 \\
\midrule
\multirow{2}{*}{Motor imagery}
  & Cho        &  4 & 0.066 & 0.000 \\
  & Lee        &  8 & 0.076 & 0.074 \\
\botrule
\end{tabular}
\end{table}
\subsection{Adversarial attack results}\label{subsec3-2}

Table~\ref{tab:adv_summary} reports the dominant UCB arm, adversarial FAR at $p=10$ under the UCB bandit, and FAR under the VAE and SGAN generators. Full results across $p \in \{5,10,15,20,25\}$ are in Supplementary Tables 7--9; per-arm FAR figures are in Supplementary Figs. 3--10.

\begin{table}[h]
\caption{Adversarial FAR at $p{=}10$ for UCB, VAE, and SGAN (mean\,$\pm$\,std across subjects, $\lambda=0.3$). ``Best arm'' is the UCB arm achieving highest dual reward; on Cho no arm achieves acceptance so the best arm is identified by budget reward only (marked $*$). All Mann-Whitney~U tests comparing best vs.\ worst arm are significant at $p<0.001$; see Supplementary Table 3.}\label{tab:adv_summary}
\begin{tabular}{@{}llccc@{}}
\toprule
Dataset & Best arm (UCB) & UCB FAR & VAE FAR & SGAN FAR \\
\midrule
Zhang       & Gauss.\ noise      & $0.135\pm0.119$ & $0.029\pm0.029$ & $0.024\pm0.033$ \\
Won         & Cov.\ interp.      & $0.162\pm0.068$ & $0.000\pm0.000$ & $0.000\pm0.000$ \\
\midrule
Wang-EO     & Chan.\ coupling    & $0.784\pm0.256$ & $0.798\pm0.240$ & $0.813\pm0.254$ \\
Wang-EC     & Chan.\ coupling    & $0.760\pm0.241$ & $0.806\pm0.205$ & $0.802\pm0.213$ \\
COG-BCI-EO  & Chan.\ coupling    & $0.857\pm0.147$ & $0.889\pm0.147$ & $0.883\pm0.142$ \\
COG-BCI-EC  & Chan.\ coupling    & $0.788\pm0.262$ & $0.824\pm0.245$ & $0.818\pm0.249$ \\
\midrule
Cho         & Corr.\ noise$^*$   & $0.000\pm0.000$ & $0.016\pm0.063$ & $0.010\pm0.042$ \\
Lee         & Gauss.\ noise      & $0.054\pm0.055$ & $0.049\pm0.127$ & $0.114\pm0.216$ \\
\botrule
\end{tabular}
\end{table}

\paragraph{Attacker lift over the zero-effort baseline.} Adversarial FAR is interpretable only relative to the acceptance an attacker obtains with no attack at all, hence we compare each value in Table~\ref{tab:adv_summary} against the zero-effort FAR, the rate at which an unmodified impostor trial is accepted against another participant's template at the same threshold, measured in the same runs (Table~\ref{tab:baseline_summary}; Supplementary Table 1). The resulting attacker lift, $\Delta\mathrm{FAR} = \mathrm{FAR}_{\mathrm{adv}} - \mathrm{FAR}_{\mathrm{zero\text{-}effort}}$, separates the paradigms differently from raw FAR, and more informatively. On RSVP the attack is the source of the acceptance ($+0.032$ on Zhang, $+0.051$ on Won, against baselines of $0.103$ and $0.111$). On motor imagery both terms are near zero. On resting-state, where raw adversarial FAR is highest, the zero-effort baseline is $0.812$--$0.896$ and lift is negative on all four streams ($-0.034$ to $-0.054$). The acceptance is already available to an impostor who replays a stolen recording untouched, and no perturbation family improves on that. The high resting-state FAR in Table~\ref{tab:adv_summary} therefore measures a paradigm that fails without an adversary, not an effective attack, and we interpret it accordingly throughout (\S\ref{sec4}).

\paragraph{Arm-preference structure.} The most informative result is not the magnitude of adversarial FAR but which arm the bandit converges to, given no paradigm information. The pattern is categorical. On all four resting-state streams the bandit immediately abandons Gaussian noise, correlated noise, and replay-with-jitter (per-arm FAR $= 0.000$) and concentrates entirely on geometric arms. Channel coupling achieves the highest dual reward on every stream, with covariance interpolation second. On RSVP, noise-based and interpolation arms dominate. Gaussian noise leads on Zhang by both per-arm FAR ($0.18$) and dual reward. Covariance interpolation leads on Won by dual reward ($0.359$), despite temporal distortion attaining a marginally higher raw per-arm FAR ($0.34$ vs.\ $0.30$) at greater perturbation cost (Supplementary Figs. 3--10). On motor imagery, no arm achieves meaningful acceptance. Arm differences are driven entirely by the budget component of the dual reward. This paradigm-specific arm ordering is stable across all tested $p$ values and $\lambda$ values (Supplementary Note 5). 

\paragraph{Generative attackers vs.\ UCB.} On resting-state, all three attackers are statistically indistinguishable ($|\mathrm{FAR}_a - \mathrm{FAR}_b| \leq 0.046$ across any pair of attackers $a,b$). The attack surface is so permissive that any optimisation strategy converges readily, and the generators do so at 5--6\% signal energy budget, below the covariance-interpolation arm. On motor imagery, all three independently confirm the paradigm-level stability (Cho: FAR$\,\approx\,0$; Lee: FAR $< 0.12$). On RSVP, the UCB decisively outperforms both generators. Won achieves zero generative acceptances, and Zhang reaches only 18--21\% of the UCB's FAR. The discrete arm selector commits to the effective perturbation family within tens of pulls. REINFORCE over a 32-dimensional latent space has not converged to a comparably targeted direction at $N_{\mathrm{iter}}=1{,}000$. This continuous score-guided optimisation of a synthetic template, in the spirit of earlier hill-climbing attacks on EEG biometrics~\cite{maiorana2013hillclimbing}, is outperformed here by a discrete search whenever the attack surface is structurally identifiable. The one exception is Lee (SGAN FAR $= 0.114$ vs.\ UCB $0.054$). Per-subject decomposition reveals a right-skewed distribution with median FAR $= 0.015$. Ten of 20 subjects achieve FAR $= 0.000$ across both sessions, while six exceed FAR $= 0.13$ (range $0.14$–$0.62$). This suggests subject-specific vulnerability rather than a population-level SGAN effect on motor imagery. Resistant subjects tend to sit farther from the population template centroid in feature space, though this does not fully account for the split. This ordering is specific to the nearest-centroid authenticator: against the linear SVM both generators reach FAR $0.19$--$0.20$ on RSVP where the UCB reaches $0.009$--$0.010$, reversing it (Supplementary Note 10).

\paragraph{Discriminator as perturbation detector.} SGAN training also yields a discriminator $D$. AUROC ranges from $0.496$ to $0.595$ across all streams (near-chance), interpreted in \S\ref{sec4} with full metrics in Supplementary Table 10.

\subsection{Classifier generalisation}\label{subsec3-3}

Table~\ref{tab:gnb_summary} reports baseline EER and adversarial FAR at $p=10$ for both classifiers side by side. The paradigm ordering is unchanged under GNB: resting-state stays highly permissive (adversarial FAR $0.58$--$0.81$), while RSVP and motor imagery collapse to near-zero ($\leq 0.001$), in fact more completely than under nearest-centroid. Critically, this is not explained by GNB being a globally weaker classifier, since on resting state, GNB's baseline EER is comparable to or better than nearest-centroid's on three of four streams (Wang-EO: $0.330$ vs.\ $0.389$; Wang-EC: $0.285$ vs.\ $0.345$; COG-BCI-EO: $0.454$ vs.\ $0.492$). The same holds for motor imagery in the other direction. GNB's baseline EER is markedly better than nearest-centroid's (Cho: $0.011$ vs.\ $0.066$; Lee: $0.018$ vs.\ $0.076$) and it is simultaneously more resistant to attack, a clean result rather than a floor effect. Won shows the same pattern (GNB EER $0.130$ vs.\ $0.153$). Two streams qualify this: on Zhang and Won, GNB's threshold calibration does not generalise as reliably as it does elsewhere. At $p=10$, held-out genuine FRR is $0.785$ (Zhang) and $0.524$ (Won), far above the nominal $0.10$ bound the calibration targets, whereas on resting-state and motor imagery it lands within $0.02$--$0.12$ of that target on every stream. This mirrors the small-enrollment calibration sensitivity documented for the linear SVM (Supplementary Note 9). Zhang and Won have the smallest enrollment sets of the eight streams (17 and 75 trials/subject), and GNB's per-feature Gaussian fit, while far less overfitting-prone than an SVM, is not immune to it at this $n$. GNB's near-zero adversarial FAR on Zhang and Won therefore reflects an operating point, that is, in practice, far stricter than the nominal $p=10$ threshold, and cannot be fully disentangled from this miscalibration rather than genuine attack resistance. Because calibration is reliable on all four resting-state streams, this caveat does not extend to the paper's central claim.

The winning attack arm under GNB on every resting-state stream is covariance interpolation or eigenvalue perturbation, the same geometric, direct-SPD-manipulation family that dominates under nearest-centroid (\S\ref{sec3}, Supplementary Figs. 3--10). Two classifiers with near-opposite structural assumptions about the feature space are exploited by the identical family of perturbations when trained on the same log-covariance representation, yielding the evidence that the vulnerability is a property of the SPD-manifold feature geometry itself, not of any one decision rule built on top of it. The pattern extends to a classifier with a learned decision boundary. A linear SVM matches or exceeds NC/GNB baseline discrimination on RSVP and motor-imagery (EER as low as $0.006$) yet shows the same near-chance discrimination and high adversarial FAR on resting-state (adversarial FAR $0.74$--$0.84$ at $p=10$), indicating the vulnerability is not attributable to classifier simplicity (Supplementary Note 9, Table 12).

\begin{table}[h]
\caption{Classifier generalisation: nearest-centroid vs.\ GNB, baseline EER and adversarial FAR at $p=10$, all eight streams. GNB best arm is the UCB arm with highest raw FAR under GNB scoring.}\label{tab:gnb_summary}
\begin{tabular}{@{}llcccc@{}}
\toprule
Dataset & GNB best arm & NC EER & NC Adv.\ FAR & GNB EER & GNB Adv.\ FAR \\
\midrule
Zhang       & Gauss.\ noise      & 0.154 & 0.135 & 0.393 & 0.000 \\
Won         & Chan.\ coupling    & 0.153 & 0.162 & 0.130 & 0.001 \\
\midrule
Wang-EO     & Eigenval.\ perturb.& 0.389 & 0.784 & 0.330 & 0.619 \\
Wang-EC     & Cov.\ interp.      & 0.345 & 0.760 & 0.285 & 0.575 \\
COG-BCI-EO  & Eigenval.\ perturb.& 0.492 & 0.857 & 0.454 & 0.811 \\
COG-BCI-EC  & Eigenval.\ perturb.& 0.391 & 0.788 & 0.400 & 0.732 \\
\midrule
Cho         & Cov.\ interp.      & 0.066 & 0.000 & 0.011 & 0.001 \\
Lee         & Cov.\ interp.      & 0.076 & 0.054 & 0.018 & 0.001 \\
\botrule
\end{tabular}
\end{table}

\subsection{Cross-condition spoofing}\label{subsec3-4}
Table~\ref{tab:crosscond} reports adversarial FAR at $p=10$ grouped by the relationship between the donor stream and the enrolled stream. Motor-imagery authenticators accept no cross-paradigm donor at all ($0.000\pm0.000$ across twelve ordered pairs), and RSVP is
only partially exposed. Resting-state authenticators accept donors recorded under an entirely different paradigm at $0.459\pm0.223$, which is $76\%$ of what same-paradigm donors obtain, and the donor's neural driver barely matters. Motor-imagery donors are accepted at $0.435$ and RSVP donors at $0.483$. Attacker lift is $-0.016$, so these are unmodified trials. An untouched Cho motor-imagery trial is accepted against a Wang-EO
template at $0.667$. The capability required to defeat a resting-state authenticator is therefore weaker than \S\ref{subsec3-2} indicates, needing no recording of the target identity, no recording of any enrolled subject, and no knowledge of the enrollment paradigm. The arm preferences of \S\ref{subsec3-2} nonetheless replicate, with eigenvalue perturbation or covariance interpolation winning on all sixteen resting-state cross-paradigm cells.

\begin{table}[h]
\caption{Cross-condition spoofing. Adversarial FAR at $p{=}10$ (mean\,$\pm$\,std across
enrolled identities), grouped by the relationship between the donor and enrolled streams.
``Same stream'' is the diagonal and serves as the within-condition reference.}
\label{tab:crosscond}
\begin{tabular}{@{}lccc@{}}
\toprule
Enrolled paradigm & Same stream & Within paradigm & Cross-paradigm \\
\midrule
RSVP           & $0.196\pm0.072$ & $0.016\pm0.011$ & $0.085\pm0.112$ \\
Resting-state  & $0.597\pm0.228$ & $0.607\pm0.206$ & $0.459\pm0.223$ \\
Motor imagery  & $0.017\pm0.017$ & $0.033\pm0.047$ & $0.000\pm0.000$ \\
\botrule
\end{tabular}
\end{table}

\section{Discussion}\label{sec4}

The central practical conclusion of this work is that resting-state EEG should not be used for biometric authentication, and the adversarial evaluation establishes this more forcefully than baseline error rates by themselves. Two facts have to be read together. First, resting-state authentication fails with no adversary present at all. The zero-effort FAR, the rate at which an unmodified impostor recording is accepted against another participant's template, is $0.812$--$0.896$ at $p=10$ (Table~\ref{tab:baseline_summary}; Supplementary Table 1) and at each stream's own EER operating point it remains $0.345$--$0.492$ (Supplementary Tables 4--6). Second, the adaptive attacker gains nothing on top of that. Attacker lift is negative on all four resting-state streams ($-0.034$ to $-0.054$), and this is not an artifact of the perturbation-budget term, since $\lambda=0$ moves resting-state FAR by at most $0.025$ (Supplementary Note 5). We therefore state the resting-state result as a claim about required attacker capability rather than about adversarial FAR: defeating a resting-state authenticator requires no attack at all, because possession of a single arbitrary resting-state recording already suffices. Cross-condition testing weakens even that requirement (\S\ref{subsec3-4}), since resting-state authenticators accept unmodified recordings of a different task, from
subjects they have never enrolled, at $0.459$. Possession of any montage-compatible EEG recording suffices, whatever the subject was doing. That our attacker adds nothing here is a property of the paradigm rather than a limitation of the attacker, since the identical bandit, arm set, and reward obtain positive lift on RSVP ($+0.032$ on Zhang, $+0.051$ on Won) and fail against motor imagery exactly where the underlying system is genuinely discriminative (EER $0.011$--$0.076$); the procedure detects vulnerability where it exists and does not manufacture it where it does not. The paradigm's failure is invariant to every axis we vary: the choice of classifier (nearest-centroid vs.\ GNB, \S\ref{subsec3-3}; linear SVM, Supplementary Note 9), the choice of enrollment session (Supplementary Note 7), and the choice of operating point (Supplementary Tables 7--9). The underlying cause is that resting-state covariance has high intra-subject trial-to-trial variability relative to inter-subject separation. Genuine and impostor cosine-similarity distributions already substantially overlap before any attack, which no threshold, classifier or template-freshness choice resolves (defence options discussed in Supplementary Note 8). RSVP and motor imagery do not share this failure mode. Motor imagery resists the full black-box and gray-box attack suite at every operating point tested (UCB FAR\,$\leq\,0.054$; the sole exception is SGAN on Lee at $0.114$, which per-subject decomposition attributes to 6 of 20 subjects rather than a population effect, \S\ref{subsec3-2}), and RSVP is attackable but tractable to defend with a challenge-response protocol (see below). 

The mechanism is interpretable at the level of SPD geometry. Stimulus-locked ERP covariance has a privileged temporal axis: the P300 component introduces a structured, time-indexed off-diagonal pattern that additive noise or temporal rearrangement can disrupt. Resting-state covariance encodes long-range functional connectivity~\cite{wang2020brainprint} with no privileged temporal axis. Additive noise is absorbed into the already-high trial-to-trial variance and produces no directed displacement toward a target template, whereas direct SPD manipulation, e.g. interpolation, eigenvalue reshaping, bypasses the variance floor entirely. Motor imagery covariance combines subject-specific focal structure ($\mu$/$\beta$ ERD over sensorimotor cortex~\cite{campisi2014brain}) with tight inter-subject separation; the resulting clusters are compact enough that even geometrically targeted arms cannot displace the query vector across an identity boundary at the perturbation magnitudes considered. This ordering tracks the geometric structure of each paradigm's covariance representation, rather than cohort size. Paradigms generating compact, well-separated per-subject SPD clusters are adversarially stable, while paradigms generating diffuse, overlapping clusters admit impostors whether or not any attack is applied. 

The generative-attacker comparison adds a methodological finding orthogonal to the paradigm result. When the attack surface is structurally identifiable (RSVP), coarse discrete search (UCB) is more query-efficient than continuous optimisation (VAE/SGAN) against a fixed-geometry decision rule, since UCB commits to the dominant family within tens of pulls while REINFORCE has not converged at $N_{\mathrm{iter}}=1{,}000$, though this ordering reverses against a learned boundary (Supplementary Note 10); when the surface is permissive (resting-state), all three attackers converge trivially regardless of algorithm. These findings indicate that resting-state's vulnerability is not merely under-explored by weaker attackers: stronger, gradient-based attackers do no better, and none of the three exceeds what an impostor obtains by replaying a stolen recording unmodified. There is nothing left to discover because the acceptance is available without an attack.

\subsection{Attack-family structure and what it exposes}\label{subsec4-1}

The bandit's arm preferences are the second substantive result of this work, and they are informative in a way that adversarial FAR alone is not. FAR reports whether a paradigm can be defeated, while the winning arm reports along which axis its accept region is open, which is the quantity a defender can act on.

The preferences are categorical rather than graded, and they are recovered without any paradigm information supplied to the attacker. On all four resting-state streams the bandit abandons Gaussian noise, correlated noise, and replay-with-jitter within tens of pulls, driving their per-arm FAR to exactly $0.000$, and concentrates on direct SPD manipulation. Channel coupling wins under nearest-centroid on every resting-state stream, and covariance interpolation or eigenvalue perturbation wins under GNB. On RSVP the ordering inverts, with noise-based and interpolation arms leading. On motor imagery no arm achieves meaningful acceptance at any operating point.

The resting-state case merits particular attention, because the arm structure shows that a permissive accept region is not a saturated one. Against a zero-effort baseline above $0.80$, most perturbation families drive acceptance to zero. What the bandit recovers is therefore which perturbations preserve an acceptance that was already available rather than which create acceptance, and the answer identifies the few directions along which the accept region is open at all. A defender reading only the aggregate FAR would conclude that the classifier accepts anything, yet it does not.

That distinction has direct defensive consequences. The winning families on resting-state share a signature that additive noise lacks. Channel coupling introduces a rank-deficient structure into the spatial covariance by construction. Eigenvalue perturbation reshapes the covariance spectrum while preserving its eigenbasis. Covariance interpolation moves the query along a geodesic between two genuine SPD points. Each is detectable in principle by statistics the authenticator already computes but currently discards, namely the eigenspectrum shape, the condition number, and the distribution of off-diagonal coupling across channel pairs relative to the enrolled population. None requires the detector to model the perturbation itself, only to notice that the presented covariance is atypically structured for a genuine trial from any enrolled subject.

We did not implement such a detector, and note that it would in any case only help where an attack is required at all, which excludes resting-state.

\subsection{Defence implications by paradigm}\label{subsec4-2}

RSVP remains attackable under the current protocol because the stimulus set is fixed across sessions, letting a single compromised recording be reused indefinitely. A challenge-response extension closes this, since each verification presents a fresh stimulus sequence with known target onsets, and acceptance requires both the identity match and P300 timing consistent with that session's targets, so an attacker's stale recording fails the freshness check. This is feasible because the enrolled template generalises across stimulus type; the same classifier achieves near-identical EER and adversarial FAR on Zhang's face task and Won's letter speller despite unrelated stimuli (\S\ref{sec3}). Protocol details and open questions in Supplementary Note 8. 

Resting-state has no comparable fix at the protocol level, since the vulnerability is a property of the feature geometry itself, not the classifier or template-freshness. Improvements would need to come from richer enrollment (longer recordings, multi-session template averaging) or paradigm-assisted designs, not perturbation-specific countermeasures (Supplementary Note 8). Given the effort required for uncertain gain, the resulting recommendation is to refrain from deploying resting-state authenticators under any threat model at all, since the capability that defeats them, possession of one arbitrary resting-state recording, sits below the weakest adversary normally considered.

Motor imagery needs no defence modification at the present scope. The natural next step is a white-box adversary with feature-pipeline access, which could in principle bypass the trial-variance floor that defeats every black-box arm tested~\cite{wu2022adversarial}. The SGAN discriminator's near-chance AUROC ($0.496$--$0.595$) is the expected minimax outcome of training $G$ against it, not evidence perturbations are undetectable in principle. A passive detector trained without simultaneous adversarial pressure is the appropriate architecture for a standalone check, left to future work (Supplementary Note 8).

\section{Conclusion}\label{sec5}

Across six datasets, two structurally distinct classifiers, a within- and cross-session protocol, and three attacker types, the clearest immediate result is that resting-state EEG is not a safe choice for biometric authentication. It is the only paradigm category where an unmodified impostor recording is accepted at a high rate ($0.812$--$0.896$ at the standard operating point) and where the adaptive attacker consequently gains nothing over that baseline, a failure that persists across classifier, enrollment session, and operating point, because the vulnerability is a property of the paradigm's covariance geometry rather than of any single implementation choice. Cross-condition testing shows that the recording does not even need to come from the enrollment paradigm (\S\ref{subsec3-4}). RSVP and motor imagery do not share this failure. Motor imagery resists the full attack suite outright, and RSVP is attackable but defensible via challenge-response. The UCB bandit recovers this paradigm-specific structure from accept/reject feedback alone, with no paradigm information provided, and a secondary methodological result, that against a fixed-geometry rule discrete search beats continuous optimisation exactly when the attack surface is structurally identifiable, holds alongside it, though the ordering reverses against a learned boundary (Supplementary Note 10). Paradigm choice hence presents itself as the first-order security parameter for EEG biometric authentication.

The central finding is that the bandit discovers paradigm-specific attack families from the reward signal alone, without any knowledge of the underlying paradigm. On resting-state EEG, noise-based arms are abandoned immediately and geometric arms (channel coupling, covariance interpolation) dominate; on RSVP, interpolation and noise-based arms lead; on motor imagery, no arm achieves meaningful acceptance at any operating point. This ordering directly reflects the geometric footprint of each paradigm's neural driver on the SPD manifold. Diffuse, overlapping covariance clusters admit impostors with or without perturbation, while compact sensorimotor clusters resist the full attack surface. 

The attacker comparison reveals a sample-efficiency reversal. Against the nearest-centroid authenticator, when the attack surface is structurally identifiable, as in RSVP, coarse discrete search over interpretable families is more query-efficient than fine-grained continuous optimisation (the ordering reverses against a learned decision boundary, Supplementary Note 10), since the UCB commits to the dominant family within tens of pulls, while the VAE and SGAN have not converged within the same 1,000-iteration budget. When the attack surface is permissive, as in resting-state, all three attackers converge trivially, and the generators do so at lower perturbation budget. Resting-state acceptance is not, however, attributable to the attack. Adversarial FAR sits at or below the zero-effort baseline on all four streams (Table~\ref{tab:baseline_summary}; Supplementary Table 1), so the paradigm's failure is a geometric limitation that precedes any adversary rather than a vulnerability our attacker introduces. Motor imagery authenticators withstand the full black-box and gray-box attack suite, with FAR$\leq 0.016$ on Cho at $p \geq 10$ across all three attackers.

Ultimately, the security in EEG biometric authentication is determined before classifier training begins, by the neural processes engaged during recording. Algorithmic sophistication alone cannot compensate for vulnerabilities introduced at the level of the recording paradigm. If EEG is to provide a reliable basis for authentication, security should therefore be understood as a property not only of the classifier, but also of the task used to elicit the biometric signal. 

\backmatter

\bmhead{Supplementary information}

Supplementary Information accompanies this paper, comprising Supplementary Notes 1--11, Supplementary Tables 1--14, and Supplementary Figures 1--10, covering a three-type attack taxonomy of threat models against EEG authenticators, of which the impersonation attacks studied here are Type~I, extended feature/classifier detail, feature extractor comparison, generative attacker architectures, $\lambda$-sensitivity results, Mann-Whitney U test results, dataset preprocessing, full baseline and adversarial-FAR breakdowns, per-arm figures, discriminator performance, cross-session evaluation, extended discussion, classifier generalisation extended to a linear SVM, the full attacker $\times$ classifier grid, and the complete cross-condition stream-pair grid.

\bmhead{Acknowledgements}

Claude (Anthropic) was used for assistance with \LaTeX{} typesetting, including mathematical notation, equation environments, and diagram construction. All scientific content, experimental design, analysis, and interpretation are entirely the authors' own.

\section*{Declarations}

\noindent\textbf{Funding.} Not applicable.

\noindent\textbf{Competing interests.} The authors declare no competing interests.

\noindent\textbf{Ethics approval and consent to participate.} Not applicable to the present authors. This study reanalyses previously published, publicly available, de-identified EEG datasets (\S\ref{subsec2-6}), each collected under the ethical approval and informed-consent procedures of its original publication~\cite{zhang_dataset,won_dataset,wang2022testretest,hinss2023cogbci,cho2017mi,lee2019mi}. No new data were collected from human participants by the present authors.

\noindent\textbf{Consent for publication.} Not applicable; no identifiable individual data or images are included.

\noindent\textbf{Data availability.} All raw EEG data analysed in this study are publicly available from their original publications~\cite{zhang_dataset,won_dataset,wang2022testretest,hinss2023cogbci,cho2017mi,lee2019mi}. Derived features, per-dataset results, and summary statistics generated in this study are available from the corresponding author upon request.

\noindent\textbf{Materials availability.} Not applicable.

\noindent\textbf{Code availability.} Code implementing the feature extraction pipeline, nearest-centroid and GNB classifiers, UCB bandit, VAE and SGAN generative attackers, and the cross-session evaluation is available from the corresponding author upon request.

\noindent\textbf{Author contribution.} P.T. conceived the study, designed and implemented the evaluation framework and all analysis code, ran the experiments, and wrote the manuscript. B.A.B. supervised the work, contributed to conceptualisation and interpretation of results through ongoing review and discussion, and reviewed the manuscript.

\includepdf[pages=-]{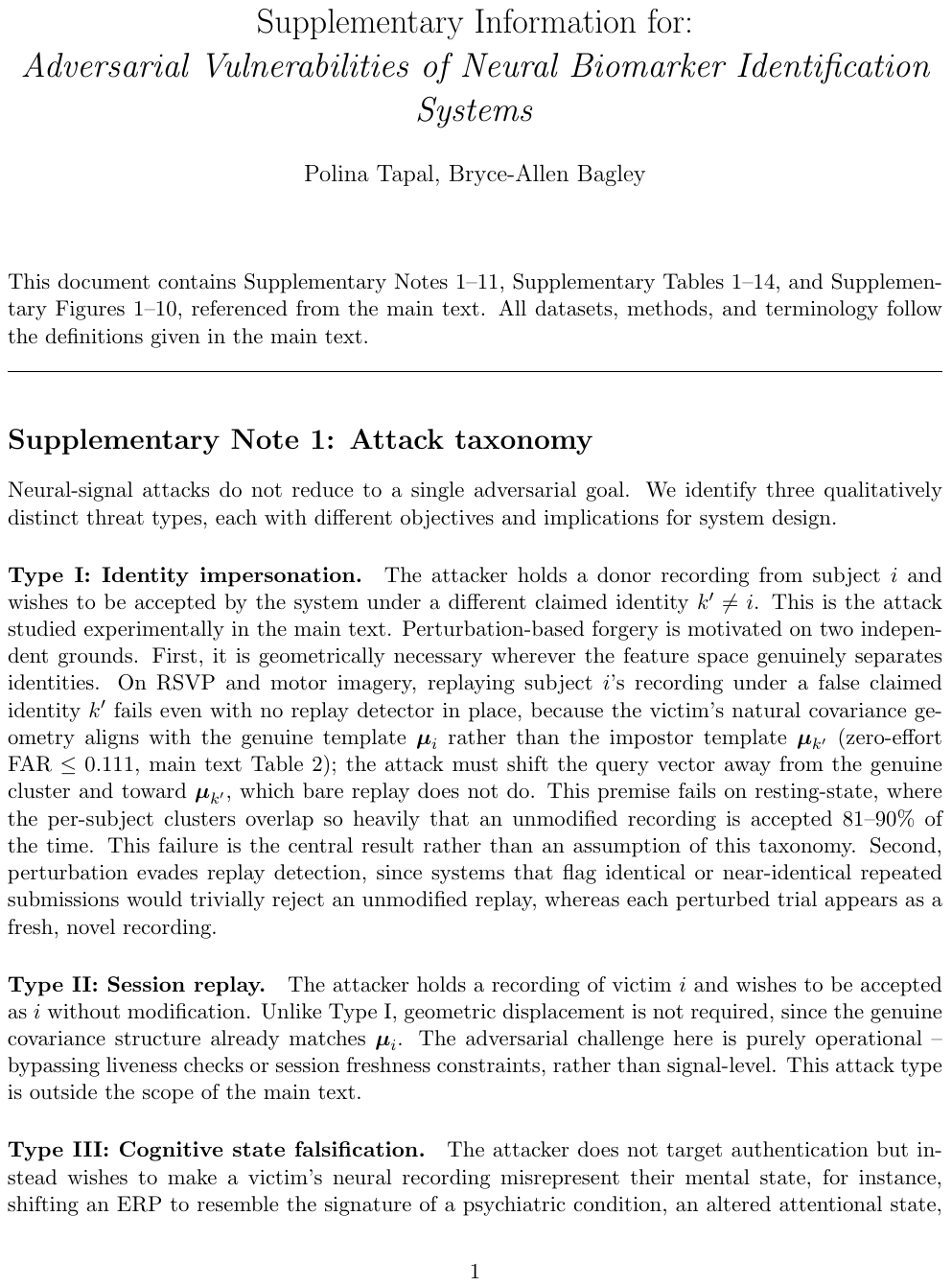}

\end{document}